\documentstyle[aaspp4,psfig,11pt]{article}
\def\ale{\mathrel{\hbox{\rlap{\hbox{\lower4pt\hbox{$\sim$}}}\hbox{$<$}}}}
\def\age{\mathrel{\hbox{\rlap{\hbox{\lower4pt\hbox{$\sim$}}}\hbox{$>$}}}}

\received{...}  \accepted{...}  \lefthead{Bloom et al.}
\righthead{Expected Characteristics of the subclass of Supernova
Gamma-ray Bursts (S-GRBs)}
\begin{document}
 
\title{\Large \bf Expected characteristics of the subclass of
Supernova Gamma-ray Bursts (S-GRBs)}
 
\author{J.~S.~Bloom\altaffilmark{1}, S.~R.~Kulkarni, F.~Harrison,
T.~Prince, E.~S.~Phinney}
\affil{California Institute of Technology, MS 105-24, Pasadena, CA 91125}
\altaffiltext{1}{email \tt{jsb@astro.caltech.edu}}

\author{and}

\author{D.~A.~Frail}
\affil{National Radio Astronomy Observatory, P.~O.~Box 0, Socorro, NM
87801}

\begin{abstract}

The spatial and temporal coincidence between the gamma-ray burst (GRB)
980425 and supernova (SN) 1998bw has prompted speculation that there
exists a class of GRBs produced by SNe (``S-GRBs''). Robust arguments
for the existence of a relativistic shock have been presented on the
basis of radio observations.  A physical model based on the radio
observations lead us to propose the following characteristics of
supernovae GRBs (S-GRBs): 1) prompt radio emission and implied
brightness temperature near or below the inverse Compton limit, 2)
high expansion velocity ($\age 50,000$ km/s) of the optical
photosphere as derived from lines widths and energy release larger
than usual, 3) no long-lived X-ray afterglow, and 4) a single pulse
(SP) GRB profile.  Radio studies of previous SNe show that only type
Ib and Ic potentially satisfy the first condition.  Accordingly we
have investigated proposed associations of GRBs and SNe finding
no convincing evidence (mainly to paucity of data) to confirm any
single connection of a SN with a GRB.  If there is a more constraining
physical basis for the burst time-history of S-GRBs beyond that of the
SP requirement, we suggest the $1 \%$ of light curves in the BATSE
catalogue similar to that of GRB 980425 may constitute the subclass.
Future optical follow-up of bursts with similar profiles should
confirm if such GRBs originate from some fraction of SN type Ib/Ic.


\end{abstract}

\section{Introduction}

With the spectroscopic observations of the optical afterglow of
gamma-ray burst (GRB) 970508 by Metzger et al.~(1997) came proof that
at least one GRB is at a cosmological distance.  Kulkarni et
al.~(1998b) later added another cosmological GRB, which, based on an
association with a high redshift galaxy, had an implied energy release
of $E_\gamma \age 10^{53}$ erg.  However, not all GRBs have been shown
to be associated with distant host galaxies.  Only about half of all
GRBs are followed by long-lived optical afterglow and one in four
produce a longer-lived radio afterglow at or above the 100 $\mu$Jy
level. In contrast, X-ray afterglow have been seen for almost all
BeppoSAX-localized bursts. Until recently, the emerging picture had
been that all GRBs are located at cosmological distances and these
GRBs (hereafter, cosmological GRBs or C-GRBs) are associated with
star-forming regions and that C-GRBs are the death-throes of massive
stars.

GRB 980425 (Galama et al.~1998) initially appeared to be one of the
those GRBs without a detectable optical (Galama et al.~1998b; Bloom et
al.~1998) or radio afterglow (Wieringa et al.~1998). However, a
possible candidate X-ray afterglow was identified (Pian et
al.~1998). Galama et al.~(1998b) noted that within the initial
localization of the GRB (as opposed to any putative X-ray afterglow
position) there was a bright optical source, later classified as SN
1998bw of type Ic. Galama et al.~(1998) argue that the probability of
finding a SNe within the 8-arcminute error circle of this GRB and an
inferred explosion date coincident within a few days of the GRB is
small, $10^{-4}$. Accordingly, they suggested that GRB 980425 is
related to the SN.

SN 1998bw showed very strong radio emission with rapid turn on and, in
fact, is the brightest radio SN to date (Wieringa et al.~1998).  This
rarity further diminished the probability of chance coincidence
(Sadler et al.~1998). Nonetheless, as with all {\it a posteriori}
estimates, it is difficult to estimate the real odds of coincidence.

From the radio observations, Kulkarni et al.~(1998c) concluded that
there exists a relativistic shock (bulk Lorentz factor, $\Gamma \equiv
(1 - \beta^2)^{-1/2} \age 2$) even 4 days after the SN explosion. This
relativistic shock appears to be slowing down as it presumably
accrets circumstellar matter. Likewise, the shock would have been
even more relativistic earlier in time and thus Kulkarni et al.~argue
that the young shock had all the necessary ingredients (high $\Gamma$,
sufficient energy) to generate a burst of $\gamma$-rays.

We feel that the physical connection between GRB 980425 and SN 1998bw
is strong. Accepting this connection then implies that there is at
least one GRB which is not of distant cosmological origin but is
instead related to a SN event in the local Universe ($\ale 100$
Mpc). We refer to this category of GRBs as supernova-GRBs or
S-GRBs. Many questions arise: How common are S-GRBs?  How can they be
distinguished from C-GRBs?  What are their typical energetics?

Attempting to answer such questions, Wang \&\ Wheeler (1998) compiled
a list of SNe that were both spatially and temporally coincident with
peculiar supernovae.  They find a strong correlation of Type Ic and Ib
SN with some GRBs. Woosley, Eastman, \& Schmidt (1998) and Iwamoto et
al. (1998) interpret GRB 980425 as a member of a new class of GRBs --
the S-GRBs. Based on their modeling of the light curve and the
spectrum of the SN they conclude that this SN was more energetic than
ordinary SNe.  The rarity once again makes the proposed association
more probable.

In this paper we accept the hypothesis that GRB 980425 is associated
with SN 1998bw. Accepting the physical model advocated by Kulkarni et
al.~(1998c) we enumerate the defining characteristics of the class of
S-GRBs. We then apply these criteria to the proposed S-GRBs (Wang \&\
Wheeler 1998, Woosley, Eastman, \& Schmidt 1998). We conclude with a
discussion of the potential number of S-GRBs.

\section{How to recognize S-GRBs?}

Below we enumerate the expected characteristics of S-GRBs. The list is
motivated by the radio observations of SN 1998bw and the model
developed by Kulkarni et al. (1998c) to explain the observations.
Briefly, from the radio data, Kulkarni et al. (1998c) conclude that
the radio emitting region is expanding at least at $2c$ (four days
after the explosion) and slowing down to $c$, one month after the
burst.  This slow down makes sense since the shock is running into
circumstellar matter and this accretion surely should slow the
shock. Applying the same logic in reverse, it is reasonable to expect
the shock to have had a higher $\Gamma$ when it was younger. The
expectation is that this high-$\Gamma$ shock is somehow responsible
for the observed burst of gamma-rays (either via synchrotron emission
or inverse-Compton scattering). In this physical model, the radio
emission and the gamma-ray emission have the same shock origin.  Of
note, whereas in C-GRBs, the primary afterglow is optical, in S-GRBs
the primary afterglow is in the radio band.

\subsection{Prompt Radio Emission and High Brightness Temperature}

An unambiguous indication of a relativistic shock in a SN is when the
inferred brightness temperature, $T_B$ exceeds $T_{\rm icc}\sim
4\times 10^{11}$ K, the so-called ``inverse Compton catastrophe''
temperature.  $T_B$ is given by (Kulkarni \&\ Phinney 1998),
$$
      T_B = 6\times 10^{8} \Gamma^{-3}\beta^{-2} S({\rm mJy})
          (\nu/5\,{\rm GHz})^{-2} t_d^{-2}d_{\rm Mpc}^2 \,\,{\rm K};
          \eqno (1)
$$
here $t_d$ is the time in days since the burst of gamma-rays, $d_{\rm
Mpc}$ is the distance in Mpc, and $S$, the flux density at frequency
$\nu$.  The energy in the particles and the magnetic field is the
smallest when $T_B \simeq T_{eq}$, the so-called ``equipartition''
temperature ($T_{eq}\sim 5\times 10^{10}$ K; Readhead 1994). The
inferred energy increases sharply with increasing $T_B$. For SN
1998bw, even with $T_B = T_{eq}$, the inferred energy in the
relativistic shock is $10^{48}$ erg which is already significant.  If
$T_B> T_{\rm icc}$ the inferred energy goes up by a factor of 500 and
thus approaches the total energy release of typical SN ($\sim 10^{51}$
erg). Thus the condition $T_B < T_{\rm icc}$ is a reasonable
inequality to use. This then leads to a lower limit on $\Gamma$.  We
consider the shock to be relativistic when $\Gamma\beta > 1$.  For SN
1998bw, Kulkarni et al.~(1998c) find $\Gamma\beta \age 2$.

It is well known that prompt radio emission (by this we mean a
timescale of a few days) is seen from Type Ib/Ic SNe. Radio emission
in Type II SNe peaks on very long timescales (months to years). No
type Ia SN has yet been detected in the radio. The reader is referred
to Weiler \&\ Sramek (1988) and Chevalier (1998) for recent
reviews. Thus the criterion of prompt radio emission (equivalent to
high $T_B$, see eq.~1) will naturally lead to selecting only type
Ib/Ic SNe.  Although slight differences may exist between type Ib/Ic
radio SNe (van Dyk et al.~1993) they can be clearly distinguished from
type II SNe not only on the basis of their prompt emission, but also
from their steeper spectral indices $\alpha$ (where
S$_\nu\propto\nu^\alpha$) and their steeper decay rates $\beta$ (where
S$_\nu\propto{t}^\beta$) (Weiler \& Sramek 1988) after the peak is
reached. Indeed, at late times SN 1998bw seems to exhibit these same
characteristics (Kulkarni et al.~1998c). van Dyk et al. (1993) first
pointed out that the 6-cm spectral luminosities of the known Type Ib/c
radio SNe showed a small scatter about $\sim$10$^{27}$ erg s$^{-1}$
Hz$^{-1}$ and thus may be considered a ``standard candle''. Thus it is
of interest to note that the peak radio luminosity of SN 1998bw is two
orders of magnitude larger than the five previously studied Type Ib/Ic
SNe.

\subsection{No Long-Lived X-ray afterglow}

In our physical picture above, we do not expect any long-lived X-ray
emission since the synchrotron lifetime of X-ray emitting electrons is
so short.  The lack of X-ray afterglow from GRB 980425 in the
direction of SN 1998bw is consistent with this picture.

\subsection{A Simple GRB profile}

In the model we have adopted, the gamma-ray and the radio emission is
powered by an energetic relativistic shock. Is it likely that there is
more than one shock? Our answer is no.  There is no basis to believe
or expect that the collapse of the progenitor core will result in
multiple shocks. It is possible that the nascent pulsar or a black
hole could be energetically important but the envelope matter surely
will dampen down rapid temporal variability of the underlying
source. From this discussion we conclude that there is only one
relativistic shock.  Thus the gamma-ray burst profile should be very
simple: a single pulse (SP). Multi-peak (MP) profiles are not a
natural expectation of the adopted physical model.  

\begin{figure}[tb]
\centerline{\psfig{file=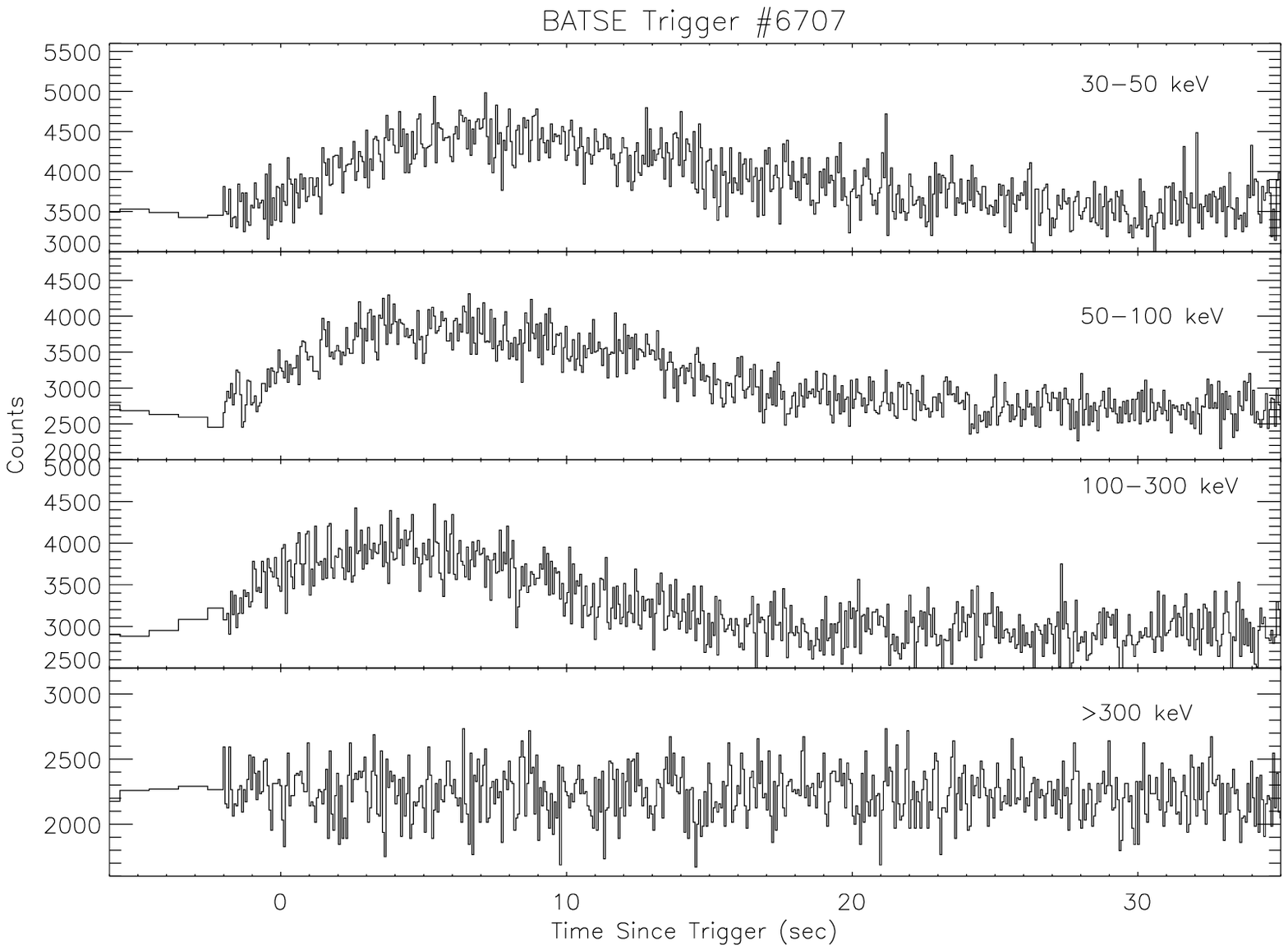,height=3in,width=5.5in}}
\centerline{\psfig{file=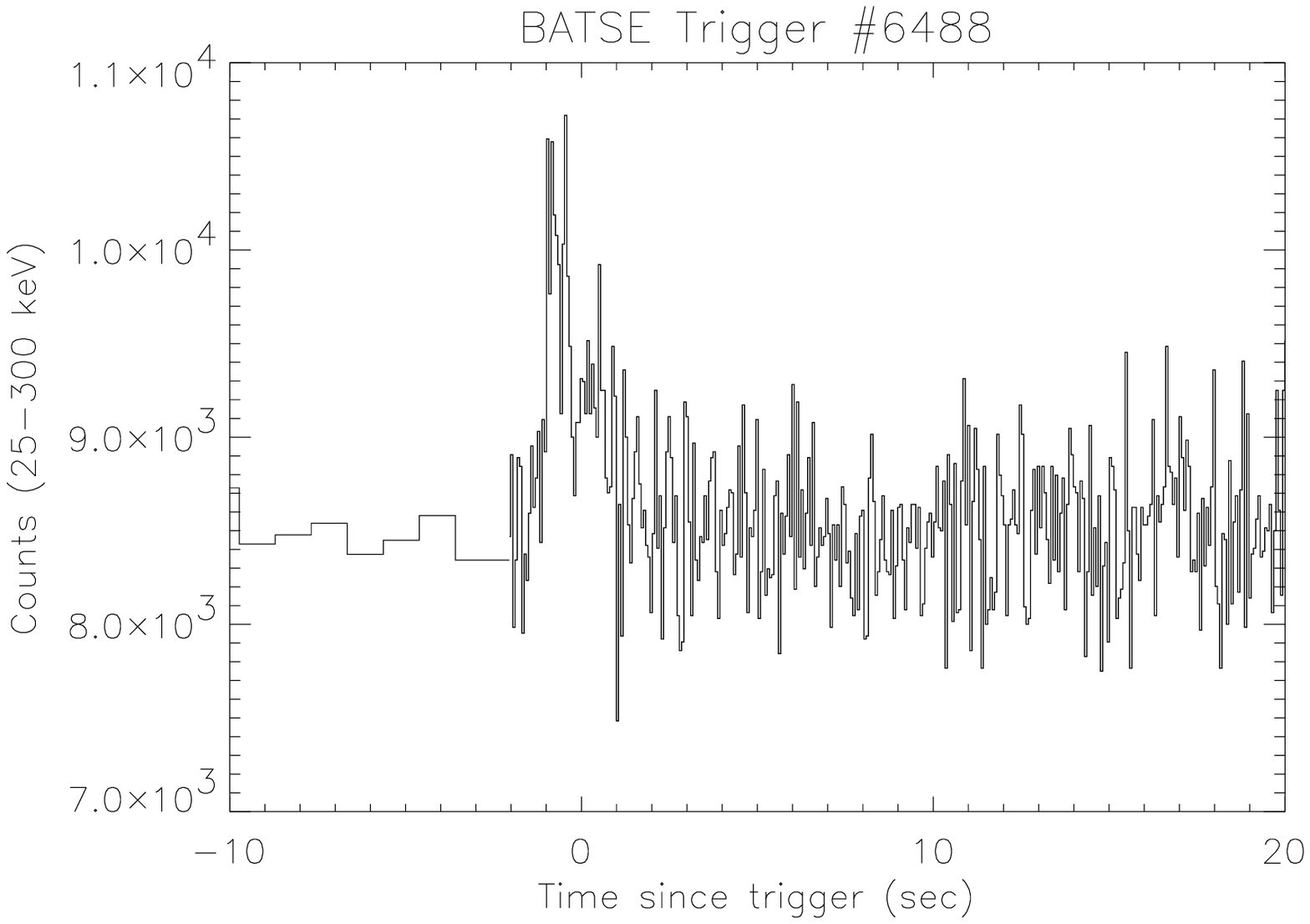,height=2in,width=2.5in}
	\psfig{file=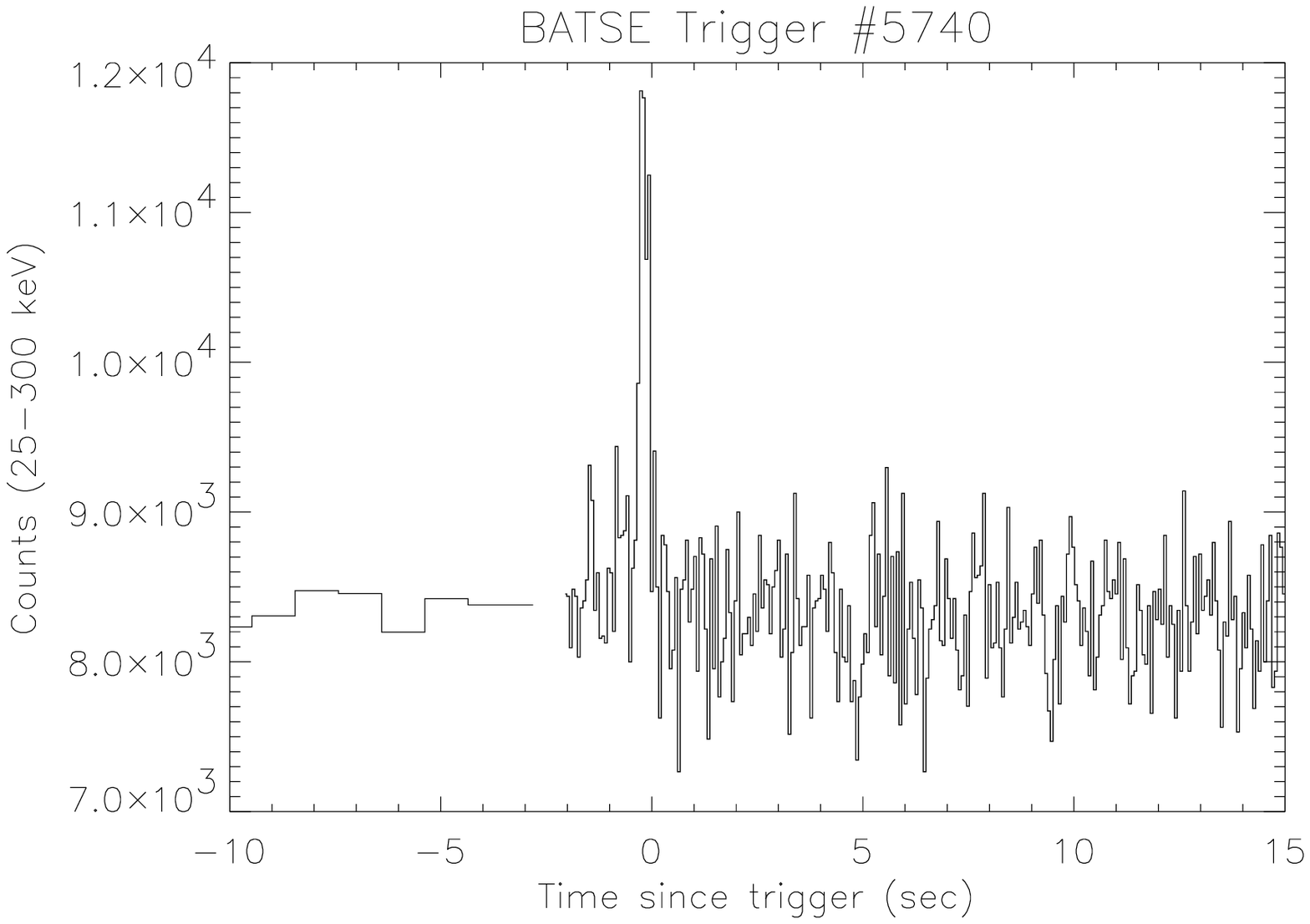,height=2in,width=2.5in}}
\centerline{\psfig{file=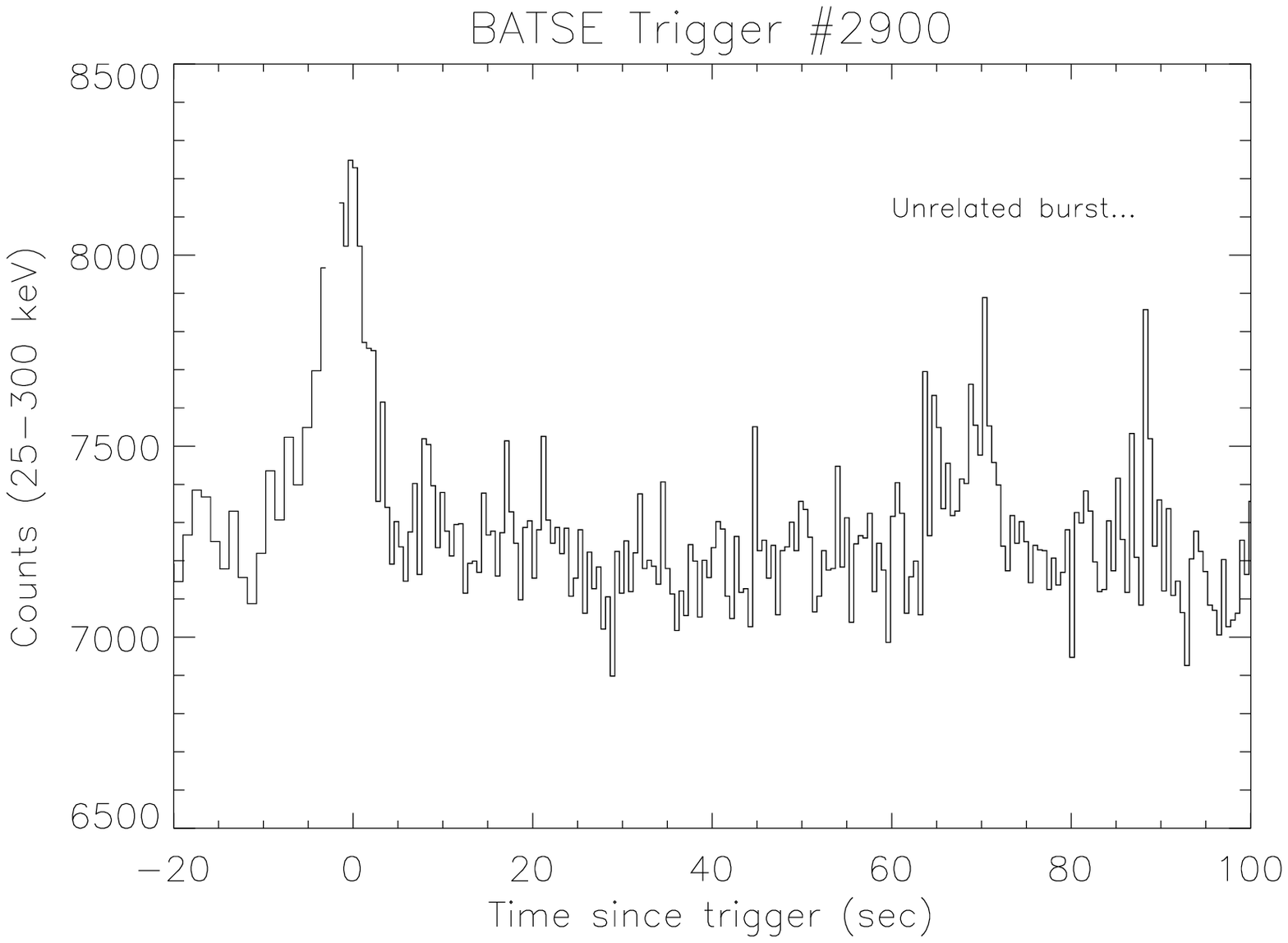,height=2in,width=2.5in}
	\psfig{file=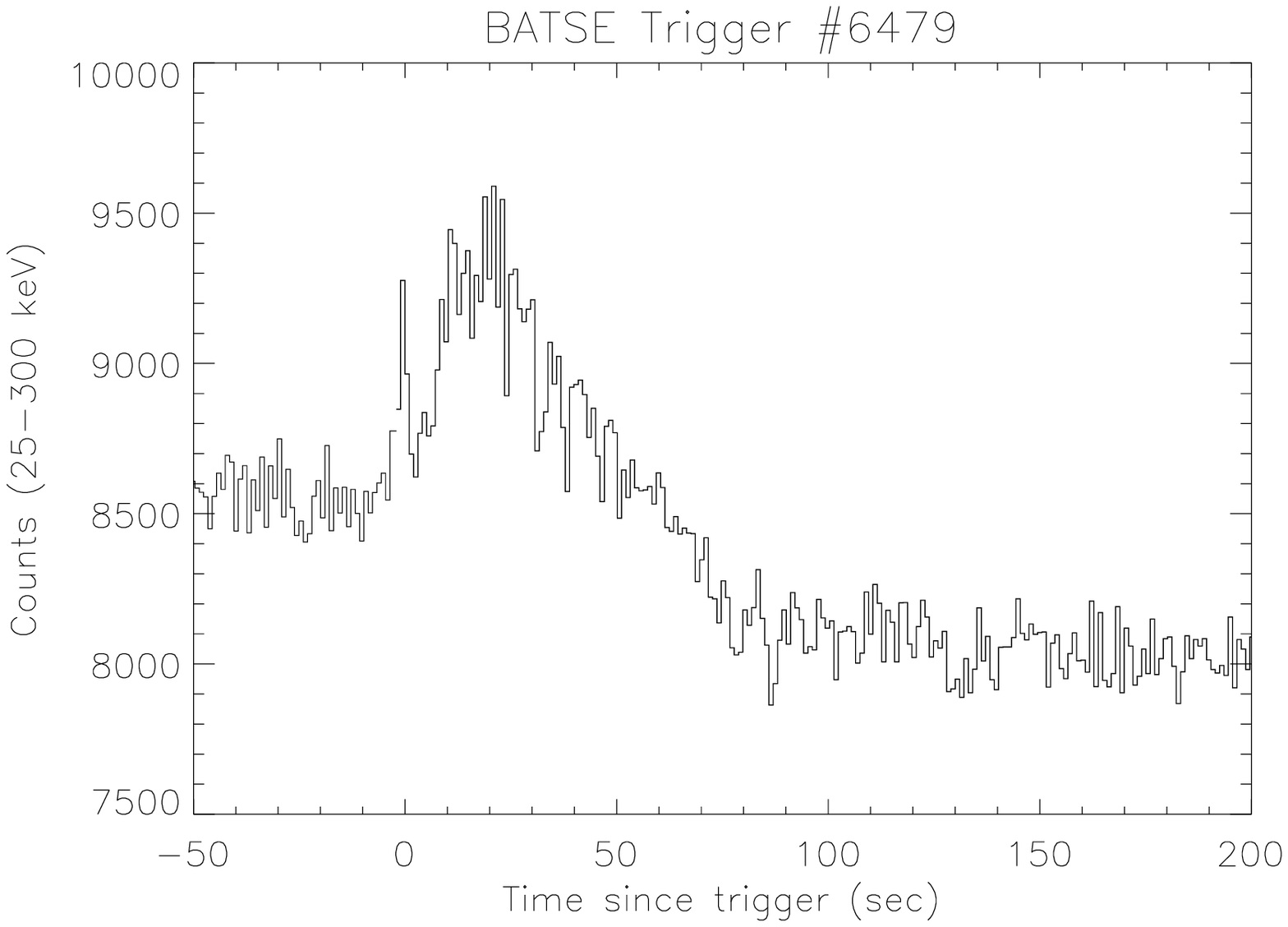,height=2in,width=2.5in}}
\caption[]{ \footnotesize (top) The 4-channel light curve of GRB
980425 (Trigger \#6707) associated with SN 1998bw. The single pulse
(SP) appears cusp-less unlike most SP BATSE bursts.  The hard-to-soft
evolution is clear from the progression of the peak from channels 3 to
1 over time.  After BATSE triggers, the light curve is sampled on
64-ms timescales. Continuous DISCLA data is augmented to the
pre-trigger light curve; this data is basically a 16 bin (1.024 sec)
averaged over the more finely sampled 64-ms data.  In the case of
longer bursts, we average the 64-ms bins over 16-sec intervals to
reduce noise.  (bottom panel; clockwise from top left) Light curve of
GRB 97112 (Trigger \#6488)/SN 1997ef or SN 1997ei; GRB 970103 (Trigger
\#5740)/SN 1997X; GRB 971115 (Trigger \#6479)/SN 1997ef; GRB 940331
(Trigger \#2900)/SN 1994I.  According to the BATSE archive, the pulse
beginning $t \simeq 65$ sec is an unrelated (ie.~not spatially
coincident) GRB.}

\label{fig:6707}
\end{figure}

The light curve of GRB 980425 (Figure \ref{fig:6707}) is a simple
single pulse (SP) with a $\sim 5$ sec rise (HWHM) and $\sim 8$ sec
decay.  Like most GRBs (Crider et al.~1997; Band 1997), the harder
emission precedes the softer emission with channel 3 (100-300 keV)
peaking $\sim 1$ sec before channel 1 (25-50 keV).  Unlike most GRB
light curves the profile of GRB 980425 has a rounded maximum instead
of a cusp.

\subsection{Broad Line Emission and Bright Optical Luminosity}

Kulkarni et al.~(1998c) noted that the minimum energy in the
relativistic shock, $E_{\rm min}$ is $10^{48}$ erg and that the true
energy content could be as high as $10^{52}$ erg. Even the lower value
is a significant fraction of energy of the total supernova release of
ordinary SNe ($E_{\rm TOT}\sim 10^{51}$ erg) especially when one
realizes that the mass within the relativistic shock ($M_R$) is
exceedingly small, $M_R \sim E_{\rm min}/\left[c^2 (\Gamma-1)\right]
\sim 10^{-6}\, M_\odot$.  It is indeed puzzling to find that the
explosion can put so much energy in such a small amount of matter.
Curiously enough, a similarly small mass is invoked to explain C-GRBs.

Clearly, a larger energy release in the supernova would favor a more
energetic shock, and hence, increase the chance such a shock could
produce a burst of $\gamma$-rays. Indeed, there are indications from
the modeling of the light curve and the spectra that the energy
release in SN 1998bw was $3\times 10^{52}$ erg (Woosley, Eastman, \&
Schmidt 1998; Iwamoto et al. 1998), a factor of $\sim 30$ larger than
the canonical SN. This then leads us to propose the final criterion:
indications of more-than-normal release of energy.  Observationally,
this release is manifested by large expansion speed which lead to the
criterion of broad emission lines and bright optical luminosity.

\section{Application of criteria to proposed associations}

We now apply the above four criteria motivated by a specific physical
model to proposed S-GRBs.  This includes associations proposed by Wang
\&\ Wheeler (1998) and Woosley, Eastman, \& Schmidt (1998).  We
searched for more potential GRB-SN associations by cross-correlating
the earlier WATCH and {\it Interplanetary Network} (IPN) localizations
(Atteia et al.~1987; Lund 1995; Hurley et al.~1997) with master
catalogue of supernovae.  We found no convincing associations in
archival GRB/SN data before the launch of the {\it Burst and Source
Transient Experiment} (BATSE).  Thus our total list remains at nine,
seven from Wang \&\ Wheeler and two from Woosley, Eastman, \& Schmidt
(1998).

We reject the following proposed associations: (1) SN 1996N/GRB 960221
(Wang \&\ Wheeler 1998).  The IPN data rule this association on
spatial grounds alone. This lack of association was independently
recognized by Kippen et al. (1998).  (2) SN 1992ar/GRB 920616 (Woosley
et al.~1998).  The associated GRB appears not to exist in the BATSE 4B
catalog (Meegan et al.~1998) and furthermore, there are no other GRBs
within a month that are spatially coincident with the SN.  (3) SN
1998T/GRB 980218 (Wang \&\ Wheeler 1998). This is ruled out on spatial
grounds from the IPN data (Kippen et al.~1998).

In Table~\ref{tab:obs} we summarize the proposed associations.  They
are ranked according to the viability of the association based on the
four criteria discussed in the previous section. The pulse profile for
each GRB is characterized as either Simple/Single Pulse (SP) or
Multipulse (MP). The standard energy criterion of ``High Energy'' and
``No High Energy'' depending on presence or absence of emission above
300 keV (Pendleton et al.~1997) is designated by HE and NHE,
respectively.  The SNe type was drawn from the literature as was the
distance to the host galaxy.  The isotropic gamma-ray energy release
is computed from the publically available fluence (BATSE 4b catalogue;
Meegan et al.~1998) and the assumed distance.

It is unfortunate that crucial information -- the early radio emission
observations -- are missing for all but one SN (1994I). SN 1994I does
have early radio emission (IAUC 5963; Rupen et al.~1994).  However,
according to Kippen et al. (1998), the associated candidate GRB 940331
is more than 4-$\sigma$ away from the location of SN 1994I.  Thus
either the GRB associated with this event is not observed by BATSE, or
this event is not a S-GRB.

\section{Discussion}

From observations (primarily radio) and analysis of SN 1998bw we have
enumerated four criteria to identify S-GRBs. We have attempted to see
how well the proposed associations of S-GRBs fare against these three
criteria. Unfortunately, we find the existing data are so sparse that
we are unable to really judge if the proposed criteria are supported
by the observations.  

The small-number of candidate associations prohibit us from drawing
any firm conclusions based on common characteristics. Nonetheless it
is of some interest to note that four of out top five candidate S-GRBs
(the exception is \#6479) are single-pulsed (SP) bursts (see figure
1). We clarify that the ordering in table 1 did not use the morphology
of the pulse profile in arriving at the rank.  We remind the reader
that roughly half of all BATSE bursts are SP and these mostly are
sharp spikes ($<$ 1 sec) or exhibit, a fast-rise followed by an
exponential decay -- the so-called ``FREDs''.

Independent of our four criteria, the expected rate of S-GRBs is
constrained by the fact that this sub-class is expected, with the
assumption of a standard candle energy release, to have a homogeneous
Euclidean ($<V/V_{\rm max}>$ = 0.5) brightness distribution.  Since
there is a significant deviation from Euclidean in the BATSE catalogue
(eg.~Fenimore et al.~1993) S-GRBs cannot comprise a majority fraction
of the BATSE catalogue.  Indeed, as studies show (eg.~Pendleton et
al.~1997), a homogeneous population, chosen phenomenologically, can
constitute at least $\sim 25\%$ of the BATSE GRB population.

From the BeppoSAX observations we know that at least 90\% of
SAX-identified GRBs have X-ray afterglow.  Thus, at least in the SAX
sample, the population of S-GRBs is further constrained to be no more
than 10\% using the criterion of no X-ray afterglow.  However, it is
well known that SAX does not trigger on short bursts, duration $\ale$
few seconds, and thus our statement only applies to the longer bursts.

Our first criterion would lead us to predict that any SN with bright
and prompt radio emission (implying a high $T_B$) as in SN 1998bw is
necessarily preceded by a burst of gamma-rays.  However, the sky is
not routinely monitored for radio SN and thus this prediction has
little practical consequence.

On the other hand, a large fraction of the sky is monitored at
gamma-ray energies and this is where our criterion of simple profiles
could conceivably be of some use.  Unfortunately, the criterion of a
single pulse is not very useful given that about half of the BATSE
bursts are SP.  As discussed above, at most $10\%$ of all bursts could
be S-GRBs.  This forces us to conclude that S-GRBs are only a certain
sub-class of SPs.  

What could be the special characteristics of this sub-class of SPs? In
search of this special sub-class we note that the profile of GRB
980425 (Figure 1) exhibits a rounded maximum and is quite
distinctive. A visual inspection of the BATSE 4B catalog shows that
there are only 15 bursts with similar profiles; we note that such
bursts constitute 1\% of the fraction of the BATSE bursts.
Interestingly, most of these bursts appear to have the same duration
as GRB 980425, although this may be due bias in our selection.

We end with some thoughts and speculation on the population of S-GRBs.
Assuming the fluence of the GRB 980425 is indicative of the subclass,
we find a canonical $\gamma$-ray energy of $E \simeq 8 \times 10^{47}
h_{65}^{-2}$ ergs.  Although BATSE triggers on flux (rather than
fluence) 80\% of bursts with fluence $S \age 8 \times 10^{-7}$ erg
cm$^{-2}$ will be detected (Bloom, Fenimore, \& in 't Zand 1997).
Thus BATSE can potentially probe the class of S-GRBs out to $\sim 100
h_{65}^{-1}$ Mpc. van den Bergh \& Tammann (1991) concluded that the
rate of Ib/Ic is roughly half that of Type II supernovae. Thus the
expected rate of type Ib/Ic supernovae is 0.3 per day out to a
distance of 100 $h_{65}^{-1}$.  This can be compared to the daily rate
of 2 -- 3 GRBs per day at the BATSE flux limit. Thus, if all type
Ib/Ic SNe produced a S-GRB then the fraction of S-GRBs is 10\%. This
is consistent with the upper limit on the fraction due to the X-ray
afterglow criterion found above.

Further progress requires observations.  Even with the poor
localization of BATSE, a Schmidt telescope equipped with large plates
can be employed to search for SNe out to a few hundred Mpc. We suggest
that such Schmidt plate searches be conducted for those bursts which
exhibit the simple pulse structure.
 
Finally, we end by noting the value of detecting {\it and} localizing
the faintest bursts. At the faint end, S-GRBs will dominate the number
counts. From this perspective, future missions should be designed to
have highest sensitivity with adequate localization.

\begin{table*}[pt]
\caption{{\bf \large GRB/Supernovae Associations}: Which are truly
S-GRBs?  \footnotesize GRB and SN properties of the suggested pairs by
Wang \& Wheeler (1998) and Woosley, Eastman, \& Schmidt (1998) are
compared against the expected criteria of S-GRBs (see \S II).  The
associations are listed in order of decreasing likelihood that the
SN/GRB falls into the S-GRB subclass. Those with the least amount of
information are placed at the bottom of the list.  The list last two
entries are SNe which are of type II and thus listed separately. }

\label{tab:obs}
\begin{tabular}{lcccccccccr}
\hline\hline 

\multicolumn{1}{l}{Association} & SN & {\small Prompt} & $\delta
\theta^\spadesuit$ & \multicolumn{1}{c}{$v_{\rm max}$} & GRB Type &
D$^{\clubsuit}$ & {\small S$^\heartsuit$ ($\times 10^{-7}$)} &
\multicolumn{1}{c}{$E^\diamondsuit$}\\
 & {\small Type} & Radio? & $(N\sigma)$ & (km s$^{-1}$) & & {\small (Mpc)} &
(erg cm$^{-2}$) & (erg) \\

\hline

1998bw/6707 & Ic & Y & 0.0 & $60,000^i$ & SP/NHE & 39.1$^f$ & 44$^g$
 & 8.1 $\times 10^{47}$\\

1997ei/6488 & Ic & NA & 2.4 & 13,000$^k$ & SP/NHE & 48.9$^a$ &
7.69 & 2.2 $\times 10^{47}$\\

1997X/5740 & Ic & NA & 3.1 & 16,000$^m$ & SP/HE & 17.0$^a$ & 5.88
& 2.0 $\times 10^{46}$\\

1994I/2900 & Ib/c & Y & 4.4 & 14,000$^j$ & SP/NHE & 7.10$^a$ &
32.6 & 2.0 $\times 10^{46}$ \\

1997ef/6488 & Ib/c? & NA & NA & 15,000$^l$ & SP/NHE & 53.8$^a$ &
7.69 & 2.7 $\times 10^{47}$\\

~~~~~~~~/6479 & Ib/c? & NA & NA & 15,000$^l$ & MP/HE & 53.8$^a$ &
99.5 & 3.4 $\times 10^{48}$\\

1992ad/1641 & Ib & NC$\dagger$ & 2.0 & NA & NA & 19.504$^a$ & NA & NA
\\

\hline

1997cy/6230 & IIPec & NA & NA & 5000$^d$ & SP/HE & 295$^d$ & 2.22 & 2
$\times 10^{48}$\\

1993J/2265 & IIt & N & NA & 13,000$^n$ & SP/NHE & 3.63$^c$ &
1.53$^c$ & 2.4 $\times 10^{44} $\\

\hline\hline
\end{tabular}
\raggedright

{\footnotesize
\noindent $^\spadesuit$ Distance of SN from BATSE position in units of
number of BATSE sigma. From Kippen et al.~(1998).  In the case of
GRB 980425, the SN 1998bw lies near the center of the small ($\sim 8$
arcmin) BeppoSAX error circle.

\noindent $^\clubsuit$ Assuming $H_0$ = 65 km s$^{-1}$ Mpc$^{-1}$ with
$D \simeq cz/H_0$ with $z$, the heliocentric redshift, from noted
reference.

\noindent $^\heartsuit$ Fluence in BATSE channels 1 -- 4 (24--1820 keV).
 From Meegan et al.~1998 (BATSE Database) unless noted.

\noindent $^\diamondsuit$ Required isotropic energy ($> 25$ keV).

\noindent $^\dagger$ The prompt radio criterion is no constrained (NC)
by the late radio detections.

\noindent $^a$ de Vaucouleurs et al.~1991. $^b$ Clocchiatti et
al.~1998.  $^c$ Freedman et al.~1984.  $^d$ Benetti, Pizzella, and
Wheatley 1997. $^e$ Nordgren et al.~1997.  $^f$ Tinney et al.~1998.
$^g$ Galama et al.~1998. $^i$ R.~A.~Stathakis communication in
Kulkarni et al.~1998c. $^i$ Wheeler et al.~(1994). $^k$ Based on a
spectrum provided in Wang, Howell, \& Wheeler (1998). $^l$ Based on
Filippenko (1997). $^m$ Benetti et al.~(1997). $^n$ Filippenko \&
Matheson (1993).
}
\end{table*}

\section*{Acknowledgments}

We thank J.~Sievers, R.~Simcoe, E.~Waxman, B.~Kirshner, A.~Filipenko,
S.~Sigurdsson, E.~E.~Fenimore for helpful discussions and direction at
various stages of this work. SRK and JSB are supported by the National
Science Foundation.

\end{document}